\documentclass[aps,superscriptaddress,twocolumn,showpacs,prl,floatfix]{revtex4} 
\usepackage{graphicx,rotating,subfigure,amsmath,amsfonts,amssymb,delarray}

\def\l{\left}
\def\r{\right}
\def\12{\frac{1}{2}}

\begin{document}
\bibliographystyle{apsrev}


\title{The dynamical spin structure factor for the anisotropic spin-$1/2$
  Heisenberg chain}


\author{R. G. Pereira}
\affiliation{Department of Physics and Astronomy, University of British
  Columbia, Vancouver, British Columbia, Canada V6T 1Z1}
\author{J. Sirker}
\affiliation{Department of Physics and Astronomy, University of British
  Columbia, Vancouver, British Columbia, Canada V6T 1Z1}
\author{J.-S. Caux}
\author{R. Hagemans}
\affiliation{Institute for Theoretical Physics, University of
  Amsterdam, 
  1018 XE Amsterdam, The Netherlands}
\author{J. M. Maillet}
\affiliation{Laboratoire de Physique, \'Ecole Normale Sup\'erieure de Lyon,
  69364 Lyon CEDEX 07, France}
\author{S. R. White}
\affiliation{Department of Physics and Astronomy, University of California, Irvine
  CA 92697, USA}
\author{I. Affleck}
\affiliation{Department of Physics and Astronomy, University of British
  Columbia, Vancouver, British Columbia, Canada V6T 1Z1}

\date{\today}

\begin{abstract}
  The longitudinal spin structure factor for the $XXZ$-chain at small
  wave-vector $q$ is obtained using Bethe Ansatz, field theory methods and the
  Density Matrix Renormalization Group. It consists of a peak with peculiar,
  non-Lorentzian shape and a high-frequency tail. We show that the width of
  the peak is proportional to $q^2$ for finite magnetic field compared to
  $q^3$ for zero field. For the tail we derive an analytic formula without any
  adjustable parameters and demonstrate that the integrability of the model
  directly affects the lineshape.
\end{abstract}

\maketitle
One of the seminal models in the field of strong correlation effects 
is the antiferromagnetic spin-$1/2$ $XXZ$-chain
\begin{equation}
H=J\sum_{j=1}^{N}\left[S_{j}^{x}S_{j+1}^{x}+S_{j}^{y}S_{j+1}^{y}+\Delta
  S_{j}^{z}S_{j+1}^{z}-hS_{j}^{z}\right]\; ,
\label{XXZ}
\end{equation}
where $J>0$ is the coupling constant
and $h$ a magnetic field. The parameter $\Delta$ describes an exchange
anisotropy and the model is critical for $-1<\Delta\leq 1$.
Recently, much interest has focused on understanding its dynamics, in
particular, the spin \cite{Zotos} 
and the heat conductivity \cite{KluemperSakai}, both at wave-vector $q=0$. A
related important question refers to dynamical correlation functions at small
but nonzero $q$, in particular the dynamical spin structure factors
$S^{\mu\mu}(q,\omega)$, $\mu=x,y,z$ \cite{Muller}. These quantities are in
principle directly accessible by inelastic neutron scattering. Furthermore,
they are important to resolve the question of ballistic versus diffusive
transport raised by recent experiments \cite{Thurber} and would also be useful
for studying Coulomb drag for two quantum wires \cite{Glazman}.  

In this letter we study the lineshape of the longitudinal structure factor
$S^{zz}(q,\omega)$ at zero temperature in the limit of small $q$. Our main
results can be summarized as follows: By calculating the form factors
$F(q,\omega)\equiv\left\langle 0\left|S_{q}^{z}\right|\alpha\right\rangle $
(here $|0\rangle$ is the ground state and $|\alpha\rangle$ an excited state)
for finite chains based on a numerical evaluation of exact Bethe Ansatz (BA)
expressions \cite{KitanineMaillet,CauxMaillet} we establish that
$S^{zz}(q,\omega)$ consists of a peak with peculiar, non-Lorentzian shape
centered at $\omega\sim vq$, where $v$ is the spin-wave velocity, and a
high-frequency tail. We find that $|F(q,\omega)|$ is a rapidly decreasing
function of the number of particles involved in the excitation. In particular,
we find for all $\Delta$ that the peak is completely dominated by two-particle
(single particle-hole) and the tail by four-particle states (denoted by 2$p$ and
4$p$ states, respectively). Including up to eight-particle as well as bound
states we verify using Density Matrix Renormalization Group (DMRG) that the
sum rules are fulfilled with high accuracy corroborating our numerical
results. By solving the BA equations for small $\Delta$ and infinite system
size analytically we show that the width of the peak scales like $q^2$ for
$h\neq 0$. Furthermore, we calculate the high-frequency tail analytically
based on a parameter-free effective bosonic Hamiltonian. We demonstrate that
our analytical results for the linewidth and the tail are in excellent
agreement with our numerical data.


For a chain of length $N$ the longitudinal dynamical structure factor is
defined by
\begin{eqnarray}
S^{zz}\left(q,\omega\right)&=&\frac{1}{N}\sum_{j,j^{\prime}=1}^{N}e^{-iq\left(j-j^{\prime}\right)}\int_{-\infty}^{+\infty}\!\!\!\!\!\!\!\!
dt\,
e^{i\omega t}\left\langle
  S_{j}^{z}\left(t\right)S_{j^{\prime}}^{z}\left(0\right)\right\rangle
\nonumber \\
&=& \frac{2\pi}{N}\sum_{\alpha}\left|\left\langle
    0\left|S_{q}^{z}\right|\alpha\right\rangle
\right|^{2}\delta\left(\omega-E_{\alpha}\right) \; .
\label{strucFac}
\end{eqnarray}
Here $S_{q}^{z}=\sum_{j}S_{j}^{z}e^{-iqj}$ and $\left|\alpha\right\rangle $ is
an eigenstate with energy $E_{\alpha}$ above the ground state energy. 
For a finite system, $S^{zz}\left(q,\omega\right)$ at fixed $q$ is a sum of
$\delta$-peaks at the energies of the eigenstates.
In the thermodynamic limit $N\rightarrow\infty$, the spectrum is continuous
and $S^{zz}\left(q,\omega\right)$ becomes a smooth function of $\omega$ and
$q$.  
By linearizing the dispersion around the Fermi points and representing the
fermionic operators in terms of bosonic ones 
the Hamiltonian (\ref{XXZ}) at low energies becomes equivalent to the Luttinger model \cite{GiamarchiBook}.
For this free boson model $S^{zz}(q,\omega)$ can be easily calculated and is given by  
\begin{equation}
S^{zz}\left(q,\omega\right)=K\left|q\right|\delta\left(\omega-v\left|q\right|\right)\, ,
\label{Szz_freeBoson}
\end{equation}
where $K$ is the Luttinger parameter. This result is a consequence of Lorentz invariance: 
a single boson with momentum $\left|q\right|$
 always carries energy $\omega=v|q|$, leading to a $\delta$-function peak 
at this level of approximation.

We expect the simple result (\ref{Szz_freeBoson}) to be modified in various
ways. First of all, the peak at $\omega\sim vq$ should acquire a finite width
$\gamma_q$. The 
latter can be easily calculated for the
$XX$ point, $\Delta=0$, where the model is equivalent to non-interacting
spinless fermions. In this case the only states that couple to the ground
state via $S^z_q$ are those containing a single particle-hole excitation
(2$p$ states). As a result, the exact $S^{zz}(q,\omega)$ is finite
only within the boundaries of the 2$p$ continuum. For $h\neq 0$, one
finds $\gamma_q\approx q^2/m$ for small $q$, where $m=(J\cos k_F)^{-1}$ is the
effective mass at the Fermi momentum $k_F$. For $h=0$, $m^{-1}\to 0$ and the
width becomes instead $\gamma_q\approx Jq^3/8$. In both cases the non-zero
linewidth is associated with the band curvature at the Fermi level and sets a
finite lifetime for the bosons in the Luttinger model.
Different attempts to calculate $\gamma_q$ for $\Delta\neq0$ have focused on
perturbation theory in the band curvature terms \cite{Samokhin} or in the
interaction $\Delta$ \cite{Kopietz,Teber} and contradictory results were found. All
these approaches have to face the breakdown of perturbation theory near
$\omega\sim vq$. 

\begin{figure}
\includegraphics*[width=0.85\columnwidth]{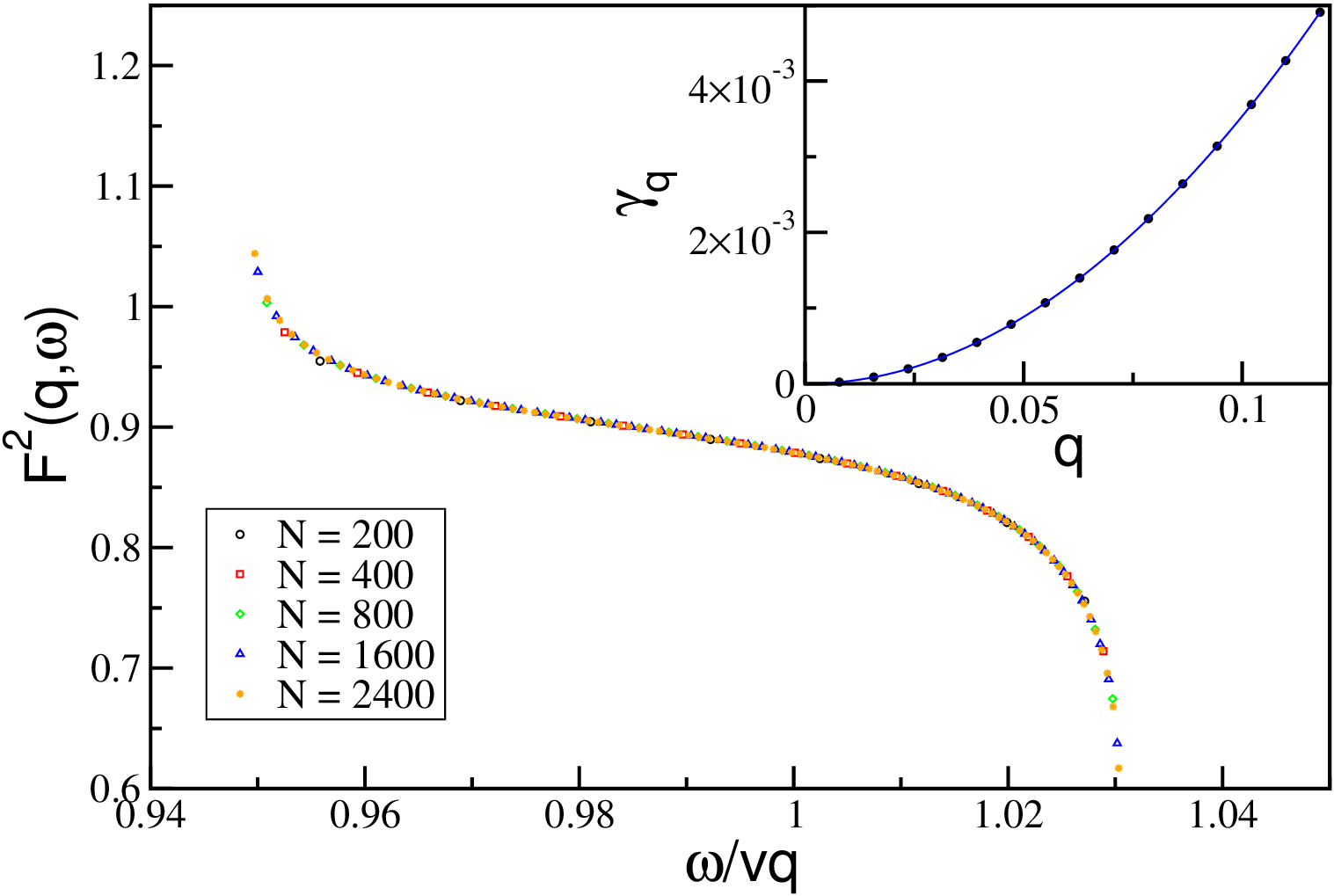}
\caption{(Color online) Form factors squared for the 2$p$ excitations and different $N$ at $\Delta=0.25$, $s=-0.1$
  and $q=2\pi/25$. 
  The inset shows the scaling of the width $\gamma_q$.
  The points are obtained by an extrapolation $N\to\infty$ of the numerical
  data. The solid line is the prediction (\ref{BA7}), $\gamma_q=0.356\, q^2$.}
\label{fig1}
\end{figure}

Since perturbative approaches show divergences on shell, our discussion about
the broadening of the peak is based on the BA solution.
The BA allows us to calculate the energy of an eigenstate exactly from a
system of coupled non-linear equations \cite{tak99}. For $\Delta=0$ these
equations decouple, the structure factor is determined by 2$p$ states
only and one recovers the free fermion solution. For $|\Delta|\ll 1$ the most
important excitations are still of the 2$p$ type and one can obtain
the energies of these eigenstates analytically in the thermodynamic limit by
expanding the BA equations in lowest order in $\Delta$.
For $h\neq0$ (\textit{i.e.}, finite magnetization $s\equiv \left\langle S_j^z\right\rangle$) this leads to 
\begin{equation}
\label{BA7}
\gamma_q = 4J \l(1+\frac{2\Delta}{\pi}\sin k_F\r)\cos k_F \sin^2\frac{q}{2}
\approx \frac{q^2}{m^*} \; .
\end{equation}
for the 2$p$ type excitations. We therefore conclude that the interaction does
not change the scaling of $\gamma_q$ compared to the free fermion case but
rather induces a renormalization of the mass given by $m\rightarrow m^* =
m/(1+2\Delta\sin k_F/\pi)$. We have verified our analytical small $\Delta$
result by calculating the form factors numerically \cite{CauxMaillet}. For all
$\Delta$, we find that excitations involving more than two particles have
negligible spectral weight in the peak region. In Fig.~\ref{fig1} we therefore
show only the form factors for the 2$p$ states and a typical set of
parameters.
The form factors, for different chain lengths, $N$, collapse onto a single
curve determining the lineshape of $S^{zz}(q,\omega)$ except for a
high-frequency tail discussed later.
The form factors are enhanced near the lower threshold $\omega_L(q)$ and
suppressed near the upper threshold $\omega_U(q)$ in contrast to the almost
flat distribution for $\Delta=0$. The lineshape agrees qualitatively with the
recent result in \cite{Glazmannew} predicting a power-law singularity at
$\omega_L(q)$ with a $q$- and $\Delta$-dependent exponent. 
The inset of Fig.~\ref{fig1} provides a numerical confirmation that
$\gamma_q\sim q^2$, with the correct pre-factor as predicted in (\ref{BA7})
for $k_F=2\pi/5$.
For zero field, the bounds of the 2$p$ continuum are known
analytically \cite{CloizeauxGaudin} and lead to a scaling $\gamma_q\sim q^3$
for 
$-1<\Delta\leq 1$. Furthermore, for $h=0$ and $\Delta=1$, an exact
result for the 2$p$ contributions to the structure factor has been
derived \cite{KarbachMueller}.


Calculating a small number of form factors for finite chains poses two
important questions: 1) For finite chains 
$S^{zz}(q,\omega)$ at small $q$ is dominated by 2$p$ excitations.  Is this
still true in the thermodynamic limit? 2) How much of the spectral weight does
the relatively small number of form factors calculated account for? We can
shed some light on these questions by considering the sum rule
$I(q)=(2\pi)^{-1}\int d\omega\, S^{zz}(q,\omega) = N^{-1}\langle S^z_q
S^z_{-q}\rangle$ where the static correlation function can be obtained by
DMRG.  As example, we consider again $\Delta=0.25$, $s=-0.1$, $q=2\pi/25$ with
$N=200$. For this case we have calculated $2,220,000$ form factors including
up to $8p$ excited states as well as bound states. Note, however, that this is
still small compared to a total number of states of $2^{200}$. In the DMRG up
to $2400$ states were kept and the ordinary two-site method was utilized but
with corrections to the density matrix to ensure good convergence with
periodic boundary conditions \cite{White}. The typical truncation error was
then $\sim 10^{-10}$ and within the accuracy of the DMRG calculation (3 parts
in $10^5$) the $2,220,000$ form factors account for 100\% of $I(q)$.
99.97\% of the spectral weight is concentrated in $I_2(q)$, the contribution caused by
the $qN/2\pi=8$ single particle-hole type excitations at $\omega\sim vq$. 
With increasing $N$ we observe an extremely slow decrease in $I_2(q)$;
however, even for a system of $2400$ sites, $I_2(q)$ is only reduced by 0.13\%
compared to the $N=200$ case. While this large $N$ behavior definitely requires
further investigation it may not be very relevant to experiments, where
effective chain lengths are limited by  defects.

Another feature missed in (\ref{Szz_freeBoson}) is the small spectral weight
extending up to high frequencies $\omega\sim J$. This is relevant in the
context of drag resistance in quantum wires because of the equivalence of
$S^{zz}(q,\omega)$ and the density-density correlation function for spinless fermions
\cite{Glazman}. To calculate the high-frequency tail we start from the
Luttinger model 
\begin{equation}
\mathcal{H}_{LL}=\frac{v}{2}\left[\Pi^2+\left(\partial_x\phi\right)^2\right] \, .
\label{LL}
\end{equation}
Here, $\phi(x)$ is a bosonic field and $\Pi(x)$ its conjugated momentum
satisfying $[\phi(x),\Pi(y)]=i\delta(x-y)$.  The slowly varying part of the
spin operator is expressed as $S^z_j\sim\sqrt{K/\pi}\, \partial_x\phi$. Note
that both $v$ and $K$ depend on $\Delta$ and $h$. In the language of the
Luttinger model, the spectral weight at high frequencies is made possible by
boson-boson interactions. Therefore, we add to the model (\ref{LL}) the
following terms
\begin{eqnarray}
 \delta\mathcal{H}(x) &=&
   \eta_{-}\left[\left(\partial_{x}\phi_{L}\right)^{3}-\left(\partial_{x}\phi_{R}\right)^{3}\right]
   \nonumber \\
 &+& \eta_{+}\left[\left(\partial_{x}\phi_{L}\right)^{2}\partial_{x}\phi_{R}-\left(\partial_{x}\phi_{R}\right)^{2}\partial_{x}\phi_{L}\right]
   \nonumber \\
 &+&\zeta_-\l[(\partial_x\phi_L)^4+(\partial_x\phi_R)^4\right]+\zeta_+(\partial_x\phi_L)^2(\partial_x\phi_R)^2
   \nonumber \\
 &+&
   \zeta_3\l[\partial_x\phi_L(\partial_x\phi_R)^3+\partial_x\phi_R(\partial_x\phi_L)^3\r]
   \nonumber \\
 &+& \lambda \cos\l(4\sqrt{\pi K}\phi+4k_Fx\r)\; ,
 \label{Ham}
 \end{eqnarray}
 where $\phi_{R,L}$ are the right- and left-moving components of the bosonic
 field with $\phi=(\phi_L-\phi_R)/\sqrt{2}$.  They obey the commutation relations
$[\partial_x\phi_{L,R}(x),\partial_x\phi_{L,R}(y)]=\mp i\partial_y\delta (x-y)$.
These are the leading irrelevant operators stemming from band curvature and the
interaction part. 
The amplitudes $\eta_\pm$, $\zeta_\pm$, $\zeta_3$ and $\lambda$ are functions
of $\Delta$ and $h$. For $h\neq 0$ the $\lambda$-term (Umklapp term) is
oscillating and can therefore be omitted at low energies. Besides, the
$\zeta$-terms have a higher scaling dimension than the $\eta_\pm$-terms, so
the latter yield the leading corrections. For $h=0$, on the other hand,
particle-hole symmetry dictates that $\eta_\pm=0$ and we must consider the
$\zeta$-terms as well as the Umklapp term. For
$\gamma_{q}\ll\omega-v\left|q\right|\ll J$ it is safe to use finite order
perturbation theory in these irrelevant terms.


In the finite field case the tail is due to the $\eta_{+}$-interaction. This
allows for intermediate states with one right- and one left-moving
boson, which together can carry small momentum but high energy $\omega\gg
v\left|q\right|$. 
It is convenient to write the structure factor defined in (\ref{strucFac}) as
$S^{zz}(q,\omega)=-2\,\mbox{Im}\,\chi^{ret}(q,\omega)$ where
$\chi^{ret}=-(K/\pi)\langle\partial_x\,\phi\partial_x\phi\rangle$ is the retarded spin-spin correlation function. 
The correction at lowest order in $\eta_+$ to the free boson result then reads
\begin{eqnarray}
&&\!\!\!\!\!\!\!\delta\chi\left(q,i\omega_{n}\right)=\left[D_{R}^{\left(0\right)}\left(q,i\omega_{n}\right)+D_{L}^{\left(0\right)}\left(q,i\omega_{n}\right)\right]^{2}\Pi_{RL}\left(q,i\omega_{n}\right)
\nonumber \\
&&\Pi_{RL}\left(q,i\omega_{n}\right)=-\frac{2K\eta_{+}^{2}}{\pi}\int\int dxd\tau
e^{-i(qx-\omega_{n}\tau)} \\
&&\quad\quad\quad\quad\quad\quad\times D_{R}^{\left(0\right)}\left(x,\tau\right)D_{L}^{\left(0\right)}\left(x,\tau\right). \nonumber
\label{fF_selfE}
\end{eqnarray}
where $D_{R,L}^{(0)}(q,i\omega_n)=\langle\partial_x
\phi_{L,R}\,\partial_x\phi_{L,R}\rangle =\pm|q|/(i\omega_n\mp v|q|)$ are the free
boson propagators for the right- and left-movers, respectively, and
$\Pi_{RL}\left(q,i\omega_{n}\right)$ is the self-energy. The tail of
$S^{zz}(q,\omega)$ for $h\neq 0$ is then given by
\begin{equation}
\delta
S^{zz}_{\eta_+}\left(q,\omega\right)=\frac{K\eta_{+}^{2}q^{4}}{v\pi}\,\frac{\theta\left(\omega-v\left|q\right|\right)}{\omega^{2}-v^{2}q^{2}}.
\label{fF_tail}
\end{equation}

For $h=0$ a connection between the integrability of the $XXZ$-model and the
parameters in the corresponding low-energy effective theory exists
\cite{Bazhanov}. The integrability is related to an infinite set of conserved
quantities where the first nontrivial one is the energy current defined by
$J_E=\int dx\, j_E(x)$ with $\partial_x\, j_E(x) = i [\mathcal{H}(x),\int dy
\mathcal{H}(y)]$ \cite{ZotosPrelovsek}.  For the Hamiltonian (\ref{Ham}) we
find
\begin{eqnarray}
j_E &=& -\frac{v}{2}\l[(\partial_x\phi_L)^2-(\partial_x\phi_R)^2\r]
- 4\zeta_-\l[(\partial_x\phi_L)^4-(\partial_x\phi_R)^4\r] \nonumber \\
&+& 2\zeta_3\l[\partial_x\phi_L(\partial_x\phi_R)^3-\partial_x\phi_R(\partial_x\phi_L)^3\r]
+\ldots
\; , 
\label{heat}
\end{eqnarray}
where the neglected terms contain more than four derivatives. 
Now conservation of the energy current, $[J_E, H] = 0$, implies $\zeta_3 =
0$ \footnote{$\zeta_3$ is also absent in the effective Hamiltonian in
  Ref.~\cite{Lukyanov}.}. 
The spectral weight at high frequencies is therefore given by the
$\zeta_{+}$ and $\lambda$-terms only. 

The perturbation theory for the $\zeta_+$-term is analogous to the one for the
$\eta_+$-term. 
Now the incoming left (right) boson can decay into one left (right) and two
right (left) bosons. This contribution is then given by
\begin{equation}
\delta
S^{zz}_{\zeta_+}\left(q,\omega\right)=\frac{K\zeta_+^2}{48\pi^2v^3}q^2(\omega^2-v^2q^2)\theta(\omega-v|q|)\;
.
\label{integer_tail}
\end{equation}
For the Umklapp term, we calculate the correlations following \cite{Schulz} and find
\begin{equation}
\delta
S^{zz}_\lambda\left(q,\omega\right)=Aq^2(\omega^2-v^2q^2)^{4K-3}\theta(\omega-v|q|)\; ,
\label{frac_tail}
\end{equation}
where $A=8\pi^2\lambda^2K^2(2v)^{3-8K}/\Gamma^2(4K)$. We remark that,
in a more general non-integrable model, the $\zeta_3$ term in Eq.~(\ref{Ham})
leads to an additional contribution to the tail which decreases with energy
and becomes large near $\omega\sim vq$.
The increasing tail found in the integrable case implies a non-monotonic
behavior of $S^{zz}(q,\omega )$. Eqs.~(\ref{fF_tail}), (\ref{integer_tail})
and (\ref{frac_tail}) are valid in the thermodynamic limit. We can extend
these results for finite systems and express them in terms of the form factors
appearing in (\ref{strucFac}).  For a given momentum $q=2\pi n/N$, the form
factors generated by integer dimension operators as in (\ref{fF_tail}) and
(\ref{integer_tail}) will then be situated at the discrete energies $\omega_l
=2\pi v l/N$ with $l=n+2,n+4,\cdots$. The form factors belonging to
(\ref{frac_tail}), on the other hand, will have energies $\omega_l =2\pi
v(l+4K)/N$ with $l=n,n+2,\cdots$.

\begin{figure}[!htp]
\includegraphics*[width=0.85\columnwidth]{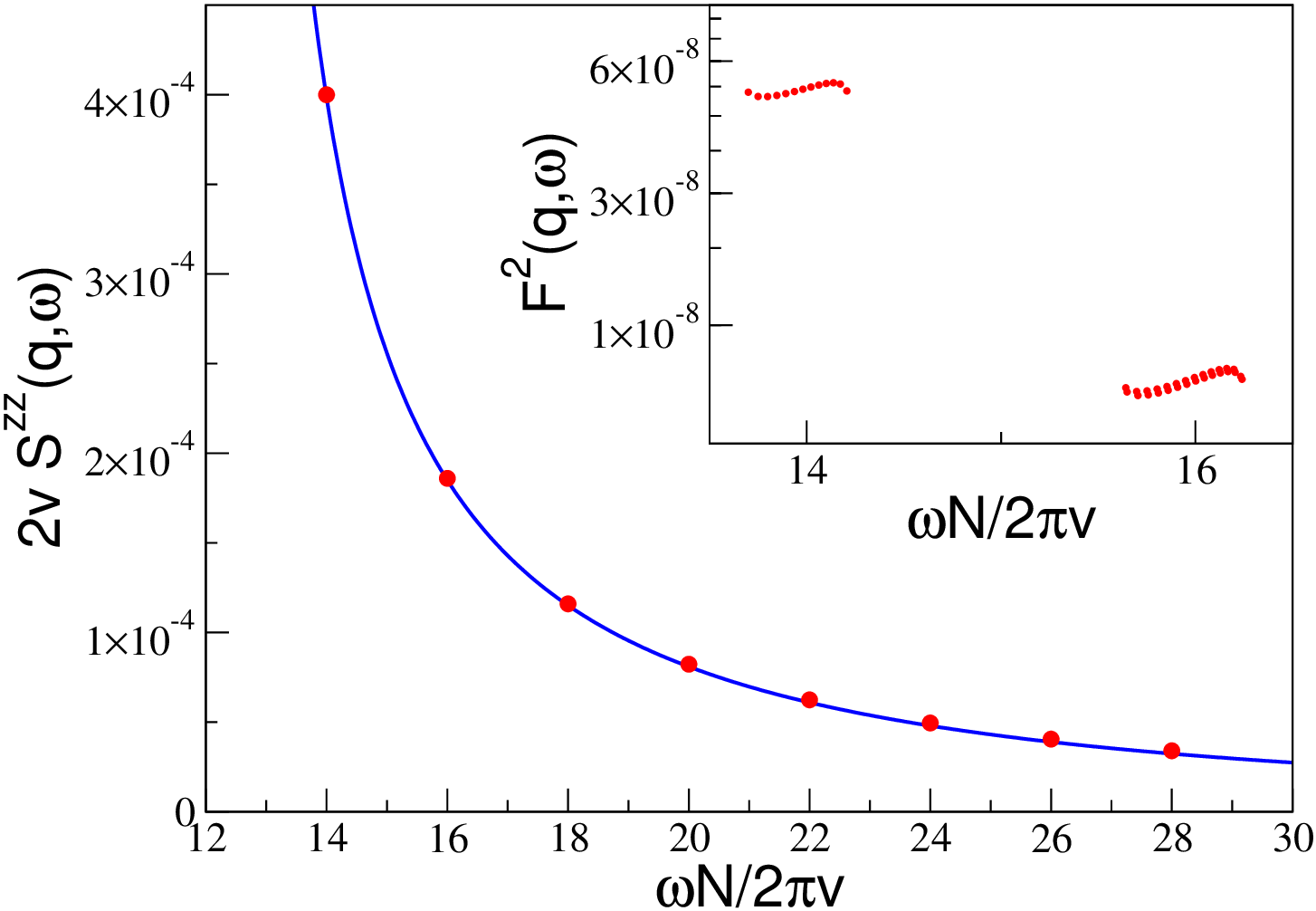}
\caption{(Color online) Form factors (dots) obtained by BA compared to
  formula (\ref{fF_tail}) (line) for $\Delta=0.75$, $s=-0.1$, $N=600$ and
  $q=2\pi/50$. The form factors for the exact eigenstates at
  $\omega\approx 2\pi v l/N$ are added and represented as a single point. The
  inset shows each form factor separately. The number of states at each level
  agrees with a simple counting based on multiple particle-hole excitations
  created around the Fermi points.}  
\label{fig2}
\end{figure}

\begin{figure}[!htp]
  \includegraphics*[width=0.85\columnwidth]{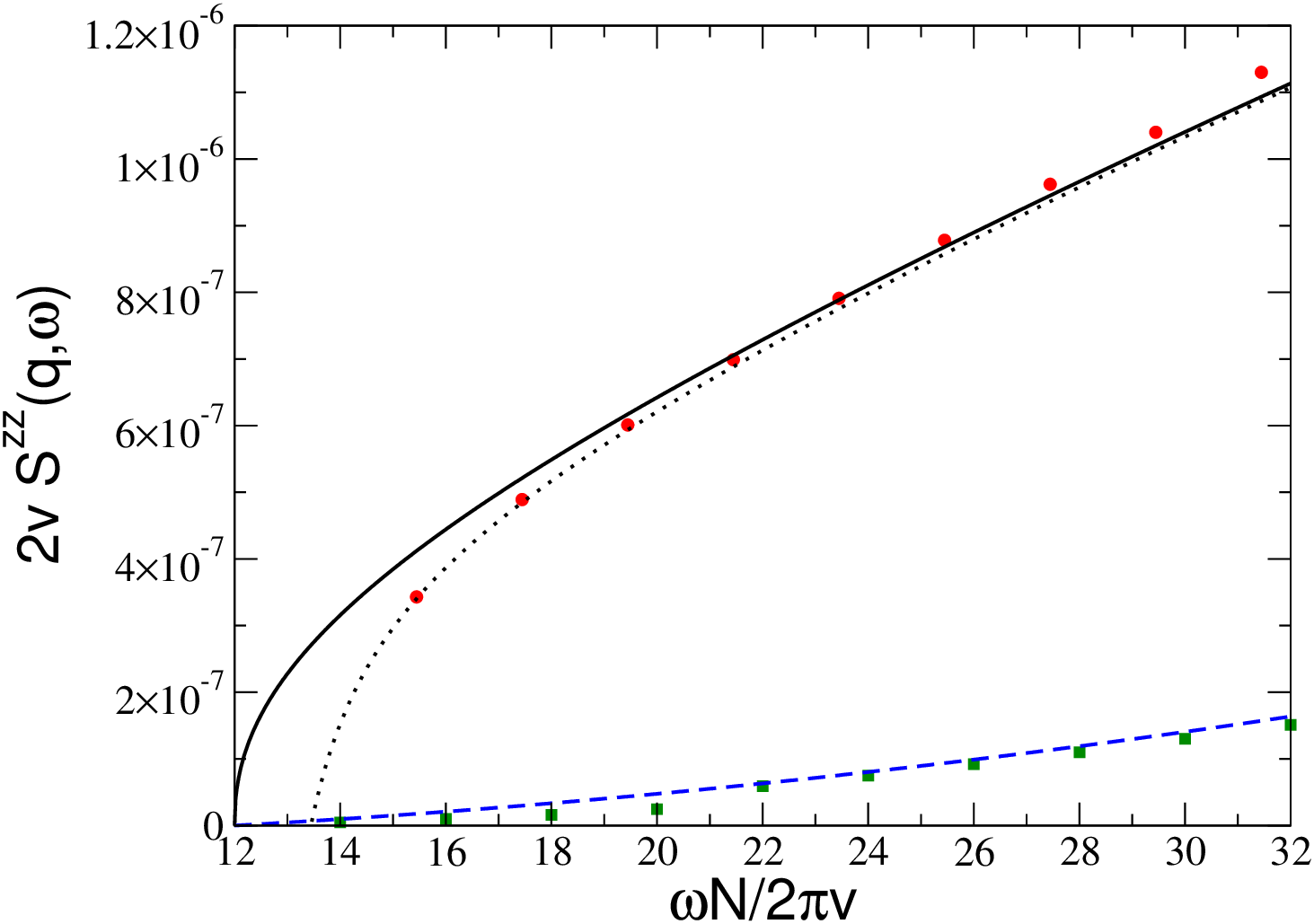}
\caption{(Color online) Sum of form factors at $\omega \approx 2\pi
  v(l+4K)/N$ (dots) and at $\omega \approx 2\pi vl/N$ (squares)
  obtained by BA for $\Delta=0.25$, $s=0$, $N=600$ and $q=2\pi/50$. The
  solid line corresponds to (\ref{frac_tail}), the dotted line to
  (\ref{frac_tail}) with finite size corrections included \cite{Note}, and the
  dashed line to (\ref{integer_tail}).}
\label{fig3}
\end{figure}

To compare our field theory results for the tail with BA data for the form
factors we have to determine the {\it a priori} unknown parameters in the effective
Hamiltonian (\ref{Ham}).  In general, they can only be obtained in terms of a
small-$\Delta$ expansion. To lowest order in $\Delta$, Eq.~(\ref{fF_tail})
reduces to the weakly interacting result in
\cite{Glazman,Teber}. 
We also checked that in this limit $\zeta_3=0$ for the $XXZ$-model but
$\zeta_3$ becomes finite if we introduce a next-nearest neighbor interaction
that breaks integrability. For an integrable model the coupling constants can
be determined by comparing thermodynamic quantities accessible by BA and field
theory. Lukyanov \cite{Lukyanov} used this procedure to find a closed form for
$\zeta_\pm$ and $\lambda$ in the case $h=0$. Similarly, the parameters
$\eta_\pm$ can be related to the change in $v$ and $K$ when varying $h$ and we
find $J\eta_-(h_0) = \sqrt{2\pi/K}v^2(a+b/2)/6$ and $J\eta_+(h_0) =
\sqrt{2\pi/K}v^2b/4$ where $a=v^{-1}\partial v/\partial h |_{h=h_0}$ and
$b=K^{-1}\partial K/\partial h |_{h=h_0}$. A numerical solution of the BA
integral equations for $v$, $K$ for infinite system size then allows us to fix
$\eta_\pm$ accurately {\it for all anisotropies and fields} so that the
formulas for the tail do not contain any free parameters. The comparison with
the form factors computed by BA for finite and zero field is shown in Figs.
\ref{fig2} and \ref{fig3}, respectively. We note that the energies of the
eigenstates are actually nondegenerate and spread around the energy levels
predicted by field theory (see inset of Fig.  \ref{fig2}).

In summary, we have presented results for the lineshape of $S^{zz}(q,\omega)$
for small $q$ based on a numerical evaluation of form factors for finite
chains. We established a linewidth $\gamma_q\sim q^2$ for $h\neq 0$ by solving
the BA equations analytically for small $\Delta$. In addition, we showed that
the spectral weight for frequencies $\gamma_{q}\ll\omega-v\left|q\right|\ll J$
is well described by the effective bosonic Hamiltonian. We presented evidence
that the lineshape of $S^{zz}(q,\omega)$ depends on the integrability of the
model. This becomes manifest in the field theory approach by a fine tuning of
coupling constants and the absence of certain irrelevant operators.



\acknowledgments We are grateful to L.I. Glazman and F.H.L. Essler for useful
discussions. This research was supported by CNPq through Grant
No.~200612/2004-2 (R.G.P), the DFG (J.S.), FOM
(J.-S.C.), CNRS and the EUCLID network (J.M.M.), the NSF under DMR 0311843
(S.R.W.), and NSERC (J.S., I.A.) and the CIAR (I.A.).

\begin{thebibliography}{18}
\bibitem[{\citenamefont{Note}(1999)}]{Note}
\bibinfo{author}{\bibfnamefont{}~\bibnamefont{Details will be published elsewhere}}.
\end{thebibliography}

\end{document}